\documentclass[conference]{IEEEtran}

\usepackage{graphicx}
\usepackage[justification=centering]{caption}
\usepackage{subfigure}
\usepackage{multirow}
\usepackage{balance}

\ifCLASSINFOpdf
\else
\fi

\usepackage{amsmath}
\usepackage{amsfonts}
\usepackage[numbers,sort&compress]{natbib}

\usepackage{algorithmic}
\usepackage[linesnumbered,boxed]{algorithm2e}

\begin{document}

\title{Hete-CF: Social-Based Collaborative Filtering Recommendation using Heterogeneous Relations}


\author{\IEEEauthorblockN{Chen Luo}
\IEEEauthorblockA{
College of Computer Science\\
and Technology\\
Jilin University, China\\
Email: rackingroll@163.com}
\and
\IEEEauthorblockN{Wei Pang}
\IEEEauthorblockA{
School of Natural and \\
Computing Sciences\\
University of Aberdeen, UK\\
Email: pang.wei@abdn.ac.uk
}
\and
\IEEEauthorblockN{Zhe Wang}
\IEEEauthorblockA{
College of Computer Science\\
and Technology\\
Jilin University, China\\
Email: wz2000@jlu.edu.cn}}

\maketitle

\begin{abstract}
Collaborative filtering algorithms haven been widely used in recommender systems. However, they often suffer from the data sparsity and cold start problems. With the increasing popularity of social media, these problems may be solved by using social-based recommendation. Social-based recommendation, as an emerging research area, uses social information to help mitigate the data sparsity and cold start problems, and it has been demonstrated that the social-based recommendation algorithms can efficiently improve the recommendation performance.
However, few of the existing algorithms have considered using multiple types of relations within one social network. In this paper, we investigate the social-based recommendation algorithms on heterogeneous social networks and proposed Hete-CF, a Social Collaborative Filtering algorithm using heterogeneous relations. Distinct from the exiting methods, Hete-CF can effectively utilize multiple types of relations in a heterogeneous social network.
In addition, Hete-CF is a general approach and can be used in arbitrary social networks, including event based social networks, location based social networks, and any other types of heterogeneous information networks associated with social information. The experimental results on two real-world data sets, DBLP (a typical heterogeneous information network) and Meetup (a typical event based social network) show the effectiveness and efficiency of our algorithm.

\end{abstract}


%
\IEEEpeerreviewmaketitle

\section{Introduction}
\label{introduction}

With the advent of Internet era and the emergence of Big Data, users constantly suffered from information overload. Therefore, recommendation systems, as effective methods to deal with information overload, become a very popular research topic in recent years \cite{p13,p4}. On the other hand, recommendation systems also help e-commerce companies provide personalised services \cite{p14}. Many companies, including \textit{YouTube} and \textit{Amazon}, have launched their own personalised recommender systems so that they can provide better services.

Among many recommendation algorithms, Collaborative Filtering \cite{p4} has been widely used in both social networks and online stores. Most of the collaborative filtering methods aim to provide recommendations or rating predictions based on historical user-item preference records \cite{p13}. However, in the real world, users only rate a limited number of items. For example, in \textit{Amazon}, a user always buys a small fraction of all available commodities, which makes the corresponding user-item information matrix very sparse. Consequently, collaborative filtering recommendation algorithms severely suffer from the cold start and data sparsity problems \cite{p26}.

In order to deal with the data sparsity problem, many algorithms have been proposed. Social-based recommendation, as one of the efficient and emerging methods, has attracted much attention in recent years \cite{p6,p1,p3,p4,p5,p13,p16}. Social-based recommendation utilises social information to help improve the recommendation performance. For example, Zhang \emph{et al.} \cite{p6} consider the recommendation system on EBSN (Event Based Social Network) \cite{p1}, which contains both online and offline networks. The algorithm presented in \cite{p6} uses the social information extracted from offline networks to help make recommendations in online networks. Another recommendation system research on EBSN also demonstrates the effectiveness of this method \cite{p3}. In \cite{p3}, the author proposed LCARS, a location-content-aware recommender system, and LCARS considers both personal interests and local preferences to make recommendations. In IJCAI-13, Yang \emph{et al.} proposed the TrustMF recommendation algorithm \cite{p4}. TrustMF considers the trust and trustee information between users from the social network. Similar to TrustMF, Xiao Yu \emph{et al.} proposed several recommendation algorithms \cite{p5,p16} based on heterogeneous information networks by introducing the relationships between items. In \cite{p13}, the cross-domain knowledge was used to improve the recommendation performance. In this paper, we will introduce the above-related research in detail in Section \ref{related}.

Previous research has demonstrated that more effective information could lead to better recommendation results \cite{p1,p4}. However, most of the above algorithms only use part of the information in social networks (either user-user or item-item information). In order to make better use of the social information, in this work we study the collaborative filtering method on heterogeneous social networks. Different from previous research, in this paper we will utilize all three types of relations, that is, not only the user-user and item-item relations, but also the user-item relations.

\begin{figure}[t]
  \centering
  \includegraphics[width=6cm]{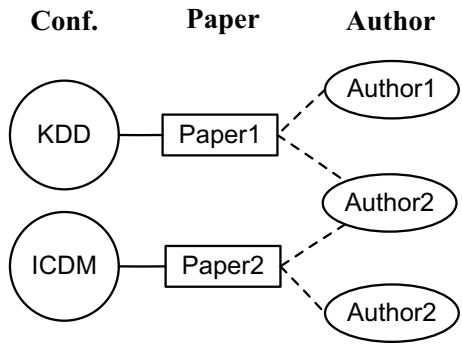}
  \centering
  \caption{A Simple Heterogeneous Network}
  \label{fig:1}
\end{figure}

As in \cite{p16,p19}, a heterogeneous social network contains more information than homogeneous ones and may have more semantic meaning. Fig.\ref{fig:1} shows a simple heterogeneous network (Bibliographic Network), and from this figure we can see that there are three types of objects in this network: Author, Paper, and Conference. It is also noted that there is more than one relation in the network.
In order to utilize these relations in a heterogeneous network, we can use meta-path \cite{p12} to present each type of relations. Meta-path is an effective way to represent relationships in heterogeneous information networks \cite{p12}. For example, the co-author relationship can be represented as a meta-path ``Author$-$Paper$-$Author"; the co-conference relationship can be represented as a meta-path ``Author$-$Paper$-$Conf.$-$Paper$-$Author".
It is pointed out that, the heterogeneous social networks in our research denotes all the HIN (heterogeneous information networks) \cite{p10} with social information, for example, EBSNs (Event-based Social Networks)\cite{p1} and LBSNs (Location-based Social Networks) \cite{p12}. We will discuss the relations between these networks in Section \ref{preliminary:HSN}.

In this research, we model the three types of relations (user-user, user-item, and item-item) respectively and propose a unified model. However, not all the social information is of benefit to the recommendation system. If the historical user-item rating is not very sparse, the social information may bias the recommendation result \cite{p13}. Therefore, we also proposed a leveraging method to evaluate the weight of the introduced social information.

Our contributions in this research are listed as follows:

1.	We proposed a general recommendation algorithm (Hete-CF) on heterogeneous social networks. The proposed algorithm can be used in many types of heterogeneous social networks, including LBSN and EBSN, etc.

2.	Three types of relations have been integrated to form a unified model, and a leveraging method has been proposed to leverage the impact of user-item relation in our model.

3.	Our approach has been tested on on two real-world data sets: DBLP, a typical heterogeneous information network with social information, and Meetup, a typical event-based social network. Experimental results demonstrate the effectiveness and efficiency of our algorithm.

The rest of the paper is organised as follows: relevant background knowledge is introduced in Section \ref{preliminary}; the proposed recommendation model is described in detail in Section \ref{Framework}. In Section \ref{Experiment}, we present our experimental results. This is followed by the introduction of some work related to our research in section \ref{related}. Finally, we conclude our research and explore future work in Section \ref{conclusion}.

\section{Preliminary}
\label{preliminary}

In this section, we first introduce some background knowledge for our research, including heterogeneous social networks \cite{p10} and meta-path \cite{p12}. Then we briefly introduce the collaborative filtering algorithm.

\subsection{Heterogeneous Social Network}
\label{preliminary:HSN}

\begin{figure} 
  \centering
  \includegraphics[width=8cm]{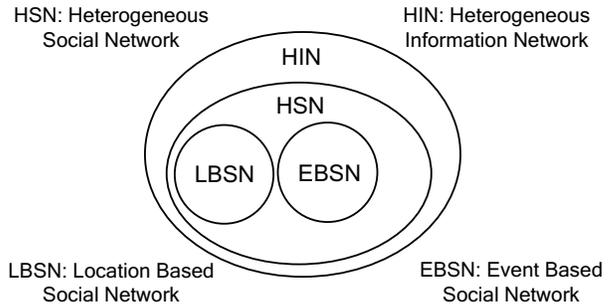}
  \centering
  \caption{Relationships between various types of Heterogeneous Networks.}
  \label{fig:2}
\end{figure}

The Heterogeneous Social Networks (HSNs) in our work can be regarded as HINs \cite{p10} containing social information, for instance, bibliographic networks, or Facebook relationship network. The relations between HSN, HIN, EBSN and LBSN are illustrated in Fig. \ref{fig:2}.

From this figure, we can see that LBSN and EBSN are special cases of HSN, and HSN is a special case of HIN. In this research, we consider all kinds of HSNs, including LBSN \cite{p11}, EBSN \cite{p1} and any other types of HIN \cite{p10} associated with social information.

As HSN is a special case of HIN, we define HSN by following the concepts of HIN. Refer to the definition of heterogeneous information networks \cite{p10}, the heterogeneous social network is defined as follows:

\newtheorem{definition}{\hskip -1.5em Definition}

\begin{definition}[\textbf{Heterogeneous Social Network}]
Suppose we have $m$ types of data objects, denoted by $X_1 = \{x_{11}, ..., x_{1n_1}\}, ..., X_m = \{x_{m1}, ..., x_{mn_m}\}$, a heterogeneous social network is in the form of a graph $G=\langle O,E,W\rangle$, where $O=\bigcup_{i=1}^{m}X_i $ ($m\geq 2$), and in $O$ there exists at least one $X_i$ which is the type of social objects (e.g., Person, Author, or Actor). $E$ is the set of links between any two data objects in $O$. $W$ is the set of weight values on the links.
\end{definition}

Obviously, $G$ will reduce to a homogeneous network when $m=1$.

In \cite{p12}, the concept of meta-path is used to denote different relations in heterogeneous information networks. In the same way as in \cite{p12}, in this research, meta-path is used to represent the relations in heterogeneous social networks. Following \cite{p12}, the meta-path is defined as follows:

\noindent
\begin{definition}[\textbf{Meta-path} \cite{p12}]
Given a heterogeneous social network $G=\langle O,E,W\rangle$, A meta path $P$ is in the form of $A_1\overset{R_1}{\rightarrow}A_2\overset{R_2}{\rightarrow}...A_{l-1}\overset{R_{l-l}}{\rightarrow}A_l $, which defines a composite relation $P=R_1\circ R_2\circ ...\circ R_{l-1}$ between two objects, and $\circ$ is the composition operator on relations.
\end{definition}

In addition, the reverse meta-path is defined as follow:

\noindent
\begin{definition}[\textbf{Reverse Meta-path} \cite{p12}]
Given a meta-path $P$: $A_l\overset{R_1}{\rightarrow}A_2\overset{R_2}{\rightarrow}...A_{l-1}\overset{R_{l-l}}{\rightarrow}A_l$, ${P}^{-1}$ is the reverse meta-path of $P$ if ${P}^{-1}$ is $A_l\overset{{R_{l-1}}^{-1}}{\rightarrow}A_{l-1}\overset{{R_{l-2}}^{-1}}{\rightarrow}...A_{2}\overset{R_l^{-1}}{\rightarrow}A_1$, where $R^{-1}$ is the reverse relation of $R$.
\end{definition}

After describing the concept of meta-path, we need to know how to measure it. Here, we use a topological measure, \textit{PathSim}, which is proposed in \cite{p12}:

Given a meta-path, denoted as $P$, the \textit{PathSim} between two objects $s$ and $t$ can be calculated as follows:

\begin{equation}
\small
S_{P}^{PS}(s,t)= \frac {2*S_{P}^{PC}(s,t)} {S_{P}^{PC}(s,:)+S_{P^{-1}}^{PC}(:,t)}
\end{equation}

In the above, $S_{P}^{PC}(s,t)$ is a \emph{Path Count} measure \cite{p12} and can be calculated as the number of path instances between $s$ and $t$. $P^{-1}$ denotes the reverse meta-path of $P$. $S_{P}^{PC}(s,:)$ denotes the path count value following $P$ and starting with $s$; and $S_{P}^{PC}(s,:)$ denotes the path count value following $P^{-1}$ and ending with $t$.

\subsection{Collaborative Filtering Model}
\label{preliminary:CF}

Collaborative filtering uses historical \textit{user-item} ratings to make recommendations. Suppose $\mathbb{R}$ is an $n-$by$-m$ matrix, which denotes the \textit{user-item} rating matrix ($n$ is the number of users and $m$ the number of items). $\mathbb{R}_{i,j}$ denotes the rating that User $i$ gives to Item $j$. One can map users and items to a lower dimension vector, denoted as $U_i$  and $V_j$, respectively. The lower dimension vectors (also called feature vectors), $U_i$  and $V_j$, can be learned by minimizing the following loss function \cite{p4}:
\begin{equation}
\small
\underset{U,V}{\min }\sum_{i=0}^{m}\sum_{j=0}^{n}(U_{i}^{T}V_{j}-\mathbb{R}_{i,j})^2+\lambda(\left \| U \right \|_F^2 + \left \| V \right \|_F^2),
\end{equation}
\noindent
where $\left \| * \right \|_F^2$ is the Frobenius norm \cite{p20}. $\lambda\in{(0,1)}$ captures the importance of each term in the loss function. $U=[U_1,U_2,...,U_n ]$ and $V=[V_1, V_2, ...,V_m]$ are the low rank representation of users and items \cite{p4}. Then the rating can be calculated as $\hat{R_{ij}} = U_i^TV_j$, where $U_{i}^{T}$ denotes the transpose of $U_{i}$. Several algorithms have been proposed to improve the performance of learning $U$ and $V$ \cite{p15,p17}. Theses algorithms will be introduced in Section \ref{related}.

\section{The Hete-CF Model}
\label{Framework}
In this section, we first propose to combine the three types of relations (user-user, user-item, and item-item) in heterogeneous social networks into a unified model. Then we introduce the leaning method of our model. At last, we introduce the complete algorithm and analyse its parameter and time complexity.

\subsection{The Recommendation Model}
\label{Framework:model}

\subsubsection{Modelling the relations between users}
\label{Framework:model:uu}

as in \cite{p4}, when we recommend items to users, the relationships between users can be used to improve the recommendation performance. This is because it is common sense that similar users may have similar orientations on a certain range of items. Let us consider an example of recommending conferences to authors for submitting their papers. Given two authors \textit{Tom} and \textit{Peter}, who focus on similar research topics and are both interested in artificial intelligence, if \textit{Tom} frequently publishes his papers in the conference \textit{AAAI}, \textit{Peter} may have a high possibility to publish his paper in \textit{AAAI} as well.

We model the above common sense by using graph regularization \cite{p9}, and the object function is shown as follows:
\begin{equation}
\small
\underset{U,A}{\min }\sum_{k=0}^{N_A}\alpha_k \sum_{i=0}^{n}\sum_{j=0}^{n} S_{A}^k(i,j)\left \| U_i-U_j \right \|_F^2
\end{equation}

In the above, $\alpha_k$ denotes the importance of the $k-$th meta-path between users and $A=[\alpha_1, \alpha_2, ..., \alpha_{N_A}]$;~$N_A$ is the number of meta-paths between users. $U_i$ has the same meaning as mentioned in Section \ref{preliminary:CF}. $S_{A}^k$ is the similarity matrix for users under the $k-$th meta-path relation, and it is calculated as follows (recall \emph{PathSim} described in Section \ref{preliminary:HSN}):
\begin{equation}
\small
S_{A}^k(i,j) = S_{P_k^{A}}^{PS}(i,j),
\end{equation}

\noindent
where $P_k^{A}$ is the $k-$th meta-path between User $i$ and User $j$.

\subsubsection{Modelling the relations between items}
\label{Framework:model:ii}

as in \cite{p15}, we can see that by introducing the relations between items the recommendation performance can be improved. It is common sense that a user may be interested in similar items. For example, if an author is interested in publishing his/her papers in \textit{ICML}, this author may also be interested in publishing his/her papers in similar conferences, for instance, \textit{NIPS} (\textit{NIPS} and \textit{ICML} are both top conferences in the field of machine learning). Here we also employ graph regularization \cite{p9} to model this common sense as follows:
\begin{equation}
\small
\underset{V,B}{\min }\sum_{k=0}^{N_B}\beta_k \sum_{i=0}^{m}\sum_{j=0}^{m} S_{B}^k(i,j)\left \| V_i-V_j \right \|_F^2
\end{equation}

In the above $\beta_k$ denotes the importance of the $k-th$ meta-path between items and $B=[\beta_1, \beta_2, ..., \beta_{N_B}]$;~$N_B$ is the number of meta-paths between items;  $V_i$ has the same meaning as mentioned in Section \ref{preliminary:CF}. $S_{B}^k$ is the similarity matrix for items; $S_{B}^k (i,j)$ is calculated as follows:
\begin{equation}
\small
S_{B}^k(i,j) = S_{P_k^{B}}^{PS}(i,j),
\end{equation}

\noindent
where $P_k^{B}$ is the $k-$th meta-path between Item $i$ and Item $j$.

\subsubsection{Modelling the relations between users and items}
\label{Framework:model:ui}

as mentioned in Section \ref{preliminary:CF}, the collaborative filtering algorithm uses historical user-item ratings to make recommendations. In a heterogeneous social network, there are also many other types of relations between users and items, and these relations may be used to further improve the recommendation performance.

For instance, if an author often cites papers from a particular conference, he may also be very likely to submit his papers to that conference. The common sense here is that users may be highly interested in the items which are ``close" to them. The term ``close" here in the context of HSNs can be represented as the larger similarity of the meta-path between a user and an item, and this distance can be calculated by using \textit{PathSim} \cite{p12}, the concept of which has been introduced in Section \ref{preliminary:HSN}. Same as the historical user-item ratings, these relationships are also between users and items. So, we can use the collaborating filtering method to model such relationships, and the model is shown below:
\begin{equation}
\small
\underset{U,V,W}\min \sum_{k=0}^{N_W}w_k\sum_{i=0}^{n}\sum_{j=0}^{m}(U_{i}^{T}V_{j}-\mathbb{R}_{i,j}^k)^2
\end{equation}

Here, $m$ and $n$ are the numbers of items and users, respectively; $w_k$ denotes the importance of the $k-$th meta-path between a user and an item, and $W=[w_1, w_2, ..., w_{N_W}]$;~$N_W$ is the number of meta-paths between users and items. $\mathbb{R}^k$ is the relation graph for the $k-$th meta-path between users and items, and it can be calculated as below:
\begin{equation}
\small
\mathbb{R}^k (i,j) = S_{P_k^{W}}^{PS}(i,j),
\end{equation}
where $P_k^{W}$ is the $k-$th meta-path between User $i$ and Item $j$.

\subsubsection{A Unified Model}
\label{Framework:model:unified}

\begin{table}\small
\caption{ The three Sub-models}
\label{submodels}
\begin{center}
\renewcommand{\arraystretch}{2}
\setlength\tabcolsep{8pt}
\begin{tabular}{c|c}
\hline
\textbf{Factor} & \textbf{Sub-model} \\
\hline
User$-$User & $\underset{U,A}{\min }\sum\limits_{k=0}^{N_A}\alpha_k \sum\limits_{i=0}^{n}\sum\limits_{j=0}^{n} S_{A}^k(i,j)\left \| U_i-U_j \right \|_F^2$ \\
\hline
Item$-$Item & $\underset{V,B}{\min }\sum\limits_{k=0}^{N_B}\beta_k \sum\limits_{i=0}^{m}\sum\limits_{j=0}^{m} S_{B}^k(i,j)\left \| V_i-V_j \right \|_F^2$ \\
\hline
User$-$Item & $\underset{U,V,W}\min \sum\limits_{k=0}^{N_W}w_k\sum\limits_{i=0}^{n}\sum\limits_{j=0}^{m}(U_{i}^{T}V_{j}-\mathbb{R}_{i,j}^k)^2$ \\
\hline
\end{tabular}
\end{center}
\end{table}

Having proposed the modelling approaches in Sections \ref{Framework:model:uu} $\sim$ \ref{Framework:model:ui}, we intend to put all the three factors together. The three corresponding sub-models are illustrated in Table \ref{submodels}. And the unified model for recommendation in a heterogeneous social network is proposed as follows:
\begin{equation}
\small
\begin{aligned}
\underset{U,V,A,B,W}{\min}
&\sum_{i=0}^{m}\sum_{j=0}^{n}(U_{i}^{T}V_{j}-\mathbb{R}_{i,j})^2\\ +
&\sum_{k=0}^{N_A}\alpha_k \sum_{i=0}^{n}\sum_{j=0}^{n} S_{A}^k(i,j)\left \|U_i-U_j \right \|_F^2\\ +
&\sum_{k=0}^{N_B}\beta_k \sum_{i=0}^{m}\sum_{j=0}^{m} S_{B}^k(i,j)\left \|V_i-V_j \right \|_F^2\\ +
&\mu\sum_{k=0}^{N_W}w_k \sum_{i=0}^{n}\sum_{j=0}^{m}(U_{i}^{T} V_{j}-\mathbb{R}_{i,j}^k)^2 \\ +
&\lambda(\left \| U \right \|_F^2 + \left \| V \right \|_F^2 +
\left \| A \right \|_F^2
+\left \| B \right \|_F^2
+\left \| W \right \|_F^2)
\end{aligned}
\end{equation}

In the above, the symbols have the same meanings as introduced in previous sections. $\mu$ and $\lambda$ are important parameters which captures the importance of each term, and we will discuss these parameters in section \ref{Framework:learning:para}.

The first term of the model incorporates the collaborating filtering component, which keeps the $U^TV$ closer to the \textit{user-item} rating matrix $\mathbb{R}$.
The second and third terms consider the \textit{user-user} and \textit{item-item} relations, respectively.
The fourth term of the model is the \textit{user-item} relationship component.  The last term of the model is the smoothing term. After minimising this model, we can obtain $U$ and $V$, and then the ratings can be obtained as $\hat{R_{ij}} = U_i^TV_j$.

In order to avoid over-fitting during the learning process, we introduce weighted-$\lambda$-regularization \cite{p15} in our algorithm. This method penalizes the feature vectors which involves more ratings. Thus the last term of our model becomes:
\begin{equation*}
\small
\lambda(\sum_i n_{user_i}\left \| U \right \|_F^2 + \sum_j n_{item_j}\left \| V \right \|_F^2 +
\left \| A \right \|_F^2
+\left \| B \right \|_F^2
+\left \| W \right \|_F^2),
\end{equation*}
where $n_{user_i}$ and $n_{item_i}$ denote the number of ratings given by User $i$ and the number of ratings given to Item $j$, respectively.

On the other hand, in a heterogeneous social network, the similarity values calculated by \emph{PathSim} are between $0$ and $1$ (recall \emph{PathSim} described in section \ref{preliminary:HSN}). So, as suggested in \cite{p17}, in order to fit the data more conveniently, we adopt a logistic function to bound the inner product of the latent feature vectors into the interval $[0,1]$. As in \cite{p17}, we use the logistic function
$f(x) = 1/{(1+{\exp}(-x))}$ in our model. Then the model to be optimized is shown below:
\begin{equation}
\small
\begin{aligned}
\underset{U,V,A,B,W}{\min}
&\sum_{i=0}^{m}\sum_{j=0}^{n}(f(U_{i}^{T}V_{j})-\mathbb{R}_{i,j})^2\\ +
&\sum_{k=0}^{N_A}\alpha_k \sum_{i=0}^{n}\sum_{j=0}^{n} S_{A}^k(i,j)\left \|U_i-U_j \right \|_F^2\\ +
&\sum_{k=0}^{N_B}\beta_k \sum_{i=0}^{m}\sum_{j=0}^{m} S_{B}^k(i,j)\left \|V_i-V_j \right \|_F^2\\ +
&\mu\sum_{k=0}^{N_W} w_k \sum_{i=0}^{n}\sum_{j=0}^{m}(f(U_{i}^{T}V_{j})-\mathbb{R}_{i,j}^k)^2 \\ +
&\lambda(\sum_i n_{user_i}\left \| U \right \|_F^2 + \sum_i n_{item_i}\left \| V \right \|_F^2 \\+
&\left \| A \right \|_F^2
+\left \| B \right \|_F^2
+\left \| W \right \|_F^2)
\end{aligned}
\end{equation}

Then the predicted rating becomes $\hat{R_{ij}}=f(U_i^TV_j)$.
However, this model cannot be directly optimised. So, in order to optimize it, we rewrite the graph regularizing terms into their trace forms as in \cite{p18}. Then the graph regularizing terms are derived as follows:
\begin{equation}
\small
\begin{aligned}
\sum_{k=0}^{N_A}\alpha_k \sum_{i=0}^{n}\sum_{j=0}^{n} S_{A}^k(i,j)\left \|U_i-U_j \right \|_F^2 \\ =
Tr(U^T(\sum_{k=0}^{N_A}\alpha_k L_{A}^k)U),
\end{aligned}
\end{equation}
and
\begin{equation}
\begin{aligned}
\sum_{k=0}^{N_B}\beta_k \sum_{i=0}^{m}\sum_{j=0}^{m} S_{B}^k(i,j)\left \|V_i-V_j \right \|_F^2 \\ =
Tr(V^T(\sum_{k=0}^{N_V}\beta_k~L_{B}^k)V).
\end{aligned}
\end{equation}

In the above, $L_{A}^k = D_A^k - S_A^k$, where $L_{A}^k$ is a diagonal matrix and $D_A^k(i,i) = \sum_{j=0}^{n}S_A^k(i,j)$.
Similarly, $L_{B}^k = D_B^k - S_B^k$, where $L_{B}^k$ is a diagonal matrix and $D_B^k(i,i) = \sum_{j=0}^{m}S_B^k(i,j)$.
Then, based on Equations (10), (11), and (12), the unified model, denoted as $J$, can be rewritten as:
\begin{equation}
\label{model}
\small
\begin{aligned}
J=
&\sum_{i=0}^{m}\sum_{j=0}^{n}(f(U_{i}^{T}V_{j})-\mathbb{R}_{i,j})^2\\ +
& Tr(U^T(\sum_{k=0}^{N_A}\alpha_k L_{A}^k)U) +
Tr(V^T(\sum_{k=0}^{N_V}\beta_k~L_{B}^k)V) \\ +
&\mu\sum_{k=0}^{N_W} w_k \sum_{i=0}^{n}\sum_{j=0}^{m}(f(U_{i}^{T}V_{j})-\mathbb{R}_{i,j}^k)^2 \\ +
&\lambda(\sum_i n_{user_i}\left \| U \right \|_F^2 + \sum_i n_{item_i}\left \| V \right \|_F^2 \\+
&\left \| A \right \|_F^2
+\left \| B \right \|_F^2
+\left \| W \right \|_F^2)
\end{aligned}
\end{equation}

\subsection{The Learning Algorithm}
\label{Framework:alg}
In this subsection, we introduce the learning algorithm of our model presented in Equation. (\ref{model}). The learning method of our model is a two-step iteration method, where the predicted rating vector $U,V$ and the weight for each meta-path $A,B,W$ mutually enhance each other. In the first step, we fix the weight vectors $A,B,W$ and learn the best predicted rating vector $U,V$. In the second step, we fix the predicted rating vector $U,V$ and learn the best weight vectors $A,B,W$.

\subsubsection{Optimize $U,V$ Given $A,B,W$}
\label{Framework:alg:uv}

When $A,B,W$ are fixed, the model becomes to a traditional Collaborative Filtering model. Therefore, as in \cite{p4}, we can use SGD (Stochastic Gradient Descent) to solve such problem. The gradient descents on $U$ are shown as follows:
\begin{equation}
\small
\begin{aligned}
\frac{1}{2}\frac{\partial J}{\partial U_i}=&\sum_{i=0}^{n}f(U_i^T V_j)^{'} U_i^T(f(U_i^TV_j)-\mathbb{R}_{i,j}) +
U_i\sum_{k=0}^{N_A}\alpha_k L_{A}^k \\+
&\mu\sum_{k=0}^{N_W}w_k \sum_{i=0}^{n}f(U_i^TV_j)^{'} U_i^T(f(U_i^TV_j)-\mathbb{R}_{i,j}^k) +
\lambda U_i,
\end{aligned}
\end{equation}
and the gradient descents on $V$ are shown as follows:
\begin{equation}
\small
\begin{aligned}
\frac{1}{2}\frac{\partial J}{\partial V_j}=&\sum_{j=0}^{m}f(U_i^TV_j)^{'} V_i(f(U_i^TV_j)-\mathbb{R}_{i,j}) + V_i\sum_{k=0}^{N_B}\beta_k L_{B}^k \\+
&\mu\sum_{k=0}^{N_W}w_k \sum_{j=0}^{m}f(U_i^TV_j)^{'} V_i(f(U_i^TV_j)-\mathbb{R}_{i,j}^k)+
\lambda V_i
\end{aligned}
\end{equation}

Then in each iteration, $U,V$ can be updated as:
\begin{equation}
\small
\begin{aligned}
U_i \leftarrow U_i - \alpha^s\frac{\partial L}{\partial U_i}~~(i=0,\cdots,n-1),
\end{aligned}
\end{equation}
and
\begin{equation}
\small
\begin{aligned}
V_i \leftarrow V_i - \alpha^s\frac{\partial L}{\partial V_j}~~(i=0,\cdots,m-1)
\end{aligned}
\end{equation}
where $\alpha^s \in (0,1)$ is used to control the convergence speed.

\subsubsection{Optimize $A,B,W$ Given $U,V$}
\label{Framework:alg:abw}
When $U,W$ is fixed, the terms only involving $U,V$ can be discarded, then the objective function can be reduced to:

\begin{equation}
\small
\begin{aligned}
J_1=
& Tr(U^T(\sum_{k=0}^{N_A}\alpha_k L_{A}^k)U) +
Tr(V^T(\sum_{k=0}^{N_V}\beta_k~L_{B}^k)V) \\ +
&\mu\sum_{k=0}^{N_W} w_k \sum_{i=0}^{n}\sum_{j=0}^{m}(f(U_{i}^{T}V_{j})-\mathbb{R}_{i,j}^k)^2 \\ +
&\lambda(
\left \| A \right \|_F^2
+\left \| B \right \|_F^2
+\left \| W \right \|_F^2)
\end{aligned}
\end{equation}

We can see that $J_1$ becomes to a linear model for each $A,B,$ and $W$. Therefore, we also use SGD (Stochastic Gradient Descent) to obtain $A,B,W$. The gradient descents on $A,B,W$ are shown as follows:
\begin{equation}
\small
\begin{aligned}
\frac{1}{2}\frac{\partial J_1}{\partial \alpha_k} =& \lambda \alpha_k
+ \sum_{i=0}^{n}\sum_{j=0}^{m}(f(U_i^T~V_j)-\mathbb{R}_{i,j})^2 \\+
& Tr(U^T(\sum_{k=0}^{N_A}\alpha_k~L_{A}^k)U)
\end{aligned}
\end{equation}
\begin{equation}
\small
\begin{aligned}
\frac{1}{2}\frac{\partial J_1}{\partial \beta_k} =& \lambda \beta_k
+ \sum_{i=0}^{n}\sum_{j=0}^{m}(f(U_i^T~V_j)-\mathbb{R}_{i,j})^2 \\+
& Tr(V^T(\sum_{k=0}^{N_B}\beta_k~L_{B}^k)V)
\end{aligned}
\end{equation}
\begin{equation}
\small
\begin{aligned}
\frac{1}{2}\frac{\partial J_1}{\partial w_k} = \lambda w_k
+ \mu\sum_{i=0}^{n}\sum_{j=0}^{m}(f(U_i^T~V_j)-\mathbb(R)_{i,j})^2
\end{aligned}
\end{equation}

Then in each iteration, $A,B,W$ can be updated as:
\begin{equation}
\small
\begin{aligned}
\alpha_i \leftarrow \alpha_i - \alpha^s \frac{\partial J_1}{\partial \alpha_i}~~(i=0,\cdots,N_A-1),
\end{aligned}
\end{equation}
and
\begin{equation}
\small
\begin{aligned}
\beta_i \leftarrow \beta_i - \alpha^s\frac{\partial J_1}{\partial \beta_j}~~(i=0,\cdots,N_B-1),
\end{aligned}
\end{equation}
and
\begin{equation}
\small
\begin{aligned}
w_i \leftarrow w_i - \alpha^s\frac{\partial J_1}{\partial w_j}~~(i=0,\cdots,N_W-1)
\end{aligned}
\end{equation}

\subsection{The Complete Algorithm}
After presenting the calculation method for each relevant variable in Section \ref{Framework:alg:uv} and Section \ref{Framework:alg:abw}, the detailed steps of our recommendation algorithm is given in Algorithm \ref{algorithm}.

\begin{algorithm}[t]
\small
\caption{Hete-CF}
\label{algorithm}
\KwIn{A heterogeneous information network $G=\langle O,E,W\rangle$.
Three sets of meta-paths between user and item, users, and items. The user-item rating matrix $\mathbb{R}$. Parameter $\lambda$, $\alpha^s$.}
\KwOut{The rating Matrix $\hat{R}$;}

Initialize $U,V,A,B,W$ randomly\;
\While {not reaching the inner $U,V,A,B,C$ difference threshold}
{
     \While {not reaching the inner $U,V$ difference threshold}
     {
        Update $U,V$ using Eqs.(16) and (17)\;
     }

    \While {not reaching the inner $A,B,C$ difference threshold}
    {
        Update $A,B,W$ using Eqs. (22), (23) and (24)\;
    }
}
The prediction rating is $\hat{R_{ij}}=f(U_i^TV_j)$.\;
return $\hat{R_{ij}}$\;
\end{algorithm}

\subsubsection{Parameter Settings}
\label{Framework:learning:para}
There are two parameters, $\lambda$ and $\mu$, in our model. As in \cite{p4,p15,p17}, $\lambda$ is always assigned manually based on the experiments and experience. Therefore, we only discuss the assignment of parameter $\mu$ in our model.

$\mu$ is an important parameter, and it can directly affect the performance of our algorithm. When $\mathbb{R}$ is sparse, a larger $\mu$ can improve the recommendation results, because more information can be added to the training process. On the other hand, when $\mathbb{R}$ is not sparse, a larger $\mu$ will bias the recommendation results.

Therefore, the value of $\mu$ depends on how sparse $\mathbb{R}$ is. In this sense, we can use the proportion of non-zero elements in matrix $\mathbb{R}$ to calculate $\mu$ as follows:
\begin{equation}
\small
\mu = \frac{\sum_{i = 1}^{n}\sum_{j = 1}^{m} I_{i,j}}{m\times n},
\label{paraset}
\end{equation}
where $I_{i,j}$ is calculated as:
\begin{equation}
\small
I_{i,j} = \begin{cases}
0 & \text{ if } \mathbb{R}_{i,j}=0 \\
1 & \text{ if } \mathbb{R}_{i,j}\neq 0
\end{cases}
\end{equation}

\subsubsection{Time Complexity Analysis}

In Algorithm \ref{algorithm}, the time for learning our model is mainly taken by computing the objective function (\ref{model}) and its corresponding gradients against feature vectors. In the inner iteration, the time complexity of computing the gradients of $J$ against $U,V$ is $O(td(n+N_W n+ N_A))$ and $O(td(m+N_W m+N_B))$, respectively. The time complexity of computing the gradients of $J$ against $W$, $A$, and $B$ is $O(td(mn))$, $O(td(mn+N_A)$, and $O(td(mn+N_B)$, respectively. In the above, $t$ is the number of iterations; $d$ is the dimensionality of feature vector. $N_W$, $N_B$, and $N_A$ have the same meaning as mentioned in Section \ref{Framework:model}, and $N_W$, $N_B$, and $N_A$ are far less than $mn$. Therefore, the upper bound of the complexity of inner iteration is $O(td(mn))$. Suppose the number of outer iteration is $T$, the overall complexity of our method is $O(Ttd(mn))$.
We can see the complexity is linearly scaling to the dimension of the \textit{user-item} numbers. The efficiency test will be introduced in Section \ref{Experiment:eff}.

\section{Experimental Results}
\label{Experiment}
In this section, we first report the experimental evaluation of our algorithm by performing a set of experiments on two real datasets. Then we test the efficiency of the algorithm on synthetic datasets. Finally, we study the parameter of our algorithm and analyse the learned weight for each selected meta-path.

\subsection{Datasets}
\label{Experiment:data}
In this research, we use two real heterogeneous social network datasets for our experiments: DBLP \footnotemark [1] and MeetUp \footnotemark [2]. As the links in DBLP and MeetUp are always sparse, we can test the ability of our algorithm on mitigating the cold start problem and data sparsity problem. On the other hand, our algorithm is proposed for the HSNs, and these two datasets are typical HSN data.

\begin{figure}[!t]
\centering
\subfigure[DBLP]
{\includegraphics[width=1.6in]{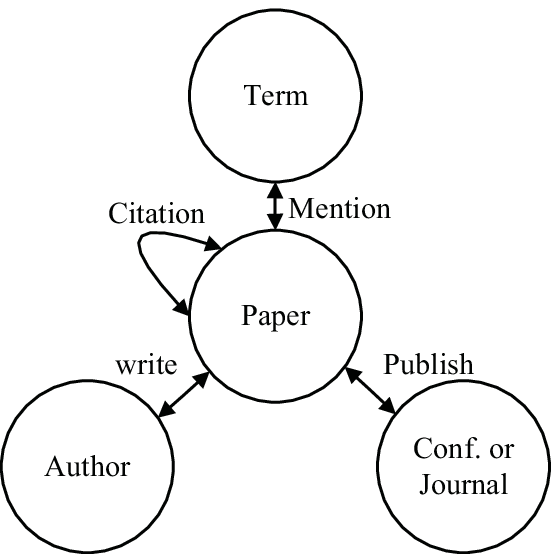}
\label{Schema:dblp}}
\subfigure[MeetUp]
{\includegraphics[width=1.6in]{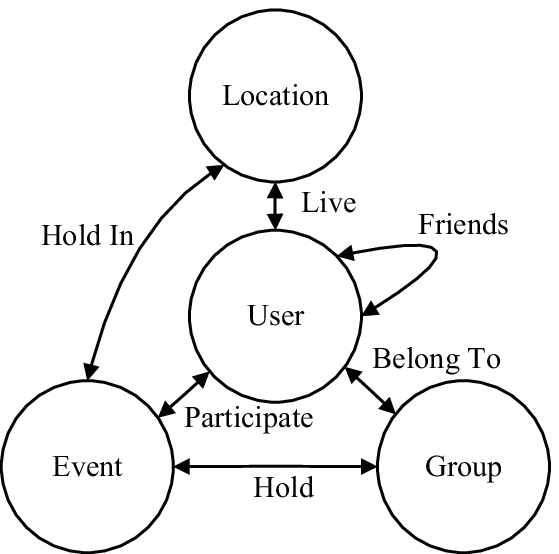}
\label{Schema:meetup}}
\caption{Network Schema}
\label{Schema}
\end{figure}

DBLP dataset is widely used for heterogeneous network analysis \cite{p5,p10,p12}. In this research, we extract a sub dataset from DBLP. The sub dataset contains the papers published in 261 computer journals and 313 computer conferences. The schema \cite{p22} of DBLP is shown in Fig.\ref{Schema:dblp}, in which Term is extracted from the titles of the papers. For the DBLP data, our recommendation problem becomes to recommend conferences to authors. Therefore, we model the historical \textit{user-item} relationship as the ``Author$-$Paper$-$Conference" meta-path, and the rate that an author gives to a conference is calculated as the \textit{PathSim} (see Section \ref{preliminary:HSN}) of the meta-path ``Author$-$Paper$-$Conference".

\footnotetext[1]{http://www.informatik.uni-trier.de/$\scriptsize{\sim}$ley/db/}
\footnotetext[2]{http://www.meetup.com/}

As in previous research on DBLP as a heterogeneous social network \cite{p10}, the meta-paths chosen for the experiments are shown in Table \ref{metapaths}. In this table, A stands for ``Author"; P stands for ``Paper", C means ``Conference or Journal", and T is for ``Term".

The second dataset is taken from Meetup, an online social website for offline meetings. The dataset comes from the one published in \cite{p1}. We extract a sub-network from the total dataset as in \cite{p6,p7,p8}. The sub-network contains the events happened in \textit{New York City} and \textit{Los Angeles} and involved the corresponding users and groups. This sub dataset is the same as the one used in \cite{p6}. The schema of Meetup network is shown in Fig. \ref{Schema:meetup}. For the Meetup dataset, we recommend groups to users. The meta-paths chosen for the experiment are shown in Table \ref{metapaths}. We have two principles for choosing the meta-paths: (1) all the types of objects need to be included into the selected meta-path; (2) choose as few meta-paths as possible. In this table, U stands for "User", L stands for "Location", G stands for ``Group", and E stands for ``Event".

\begin{table*}\small
\caption{ Meta paths selected in DBLP and MeetUp}
\label{metapaths}
\begin{center}
\renewcommand{\arraystretch}{1.1}
\setlength\tabcolsep{10pt}
\begin{tabular}{l|l|l}
\hline
\textbf{Dataset} & \textbf{Relations} & \textbf{Meta-paths}\\
\hline
\multirow{3}*{\textbf{DBLP Dataset}}
&Authors$-$Author & $A-P-A,~A-P-C-P-A,~A-P-T-P-A$ \\
\cline{2-3}
~&Conf. $-$ Conf.  & $C-P-A-P-C,~C-P-P-C,~C-P-T-P-C$ \\
\cline{2-3}
~&Author $-$ Conf.  & $A-P-T-P-C,~A-P-P-C$ \\
\hline

\multirow{3}*{\textbf{MeetUp Dataset}}
&Users $-$ User & $U-L-U$,~$U-G-U$,~$U-E-U$,~$U-U$ \\
\cline{2-3}
~&Group $-$ Group & $G-U-G,~G-E-L-E-G,~G-U-U-G$ \\
\cline{2-3}
~&Users $-$ Group & $U-U-G,~U-E-G,~U-L-E-G$ \\
\hline
\end{tabular}
\end{center}
\end{table*}

\subsection{Experimental Setup}
\label{Experiment:setup}

\subsubsection{Cross-validation}

As in \cite{p4}, we use 5-fold cross-validation for learning and testing. We randomly select $40\%$ $(60\%)$ of the data as the training set and the rest $60\%$ $(40\%)$ as the testing set. Each result discussed below is averaged over ten trials.

\subsubsection{Comparison Methods}

There are three baselines and two state-of-the-art methods used in our methods for comparison. The three baselines are listed as follows:
\noindent
\begin{itemize}
\setlength{\itemindent}{-1.5em}
\item\textbf{UserMean:} prediction rate equals the mean value of the users.
\item\textbf{ItemMean:} prediction rate equals the mean value of the items.
\item\textbf{NMF:} non-negative matrix factorization function \cite{p4}, with $d=5$, and $d=10$, where $d$ denotes the dimension of the feature vector.
\end{itemize}

The three baselines cannot make use of the heterogeneous relations in the heterogeneous social network, and they only consider the target recommendation relation (The $A-P-C$ relation in DBLP, and the $U-G$ relation in MeetUp). So in this experiment, we only consider the target recommendation relations in each dataset for these three baseline algorithms.

The two state-of-the-art methods selected in our experiment are described below:

\begin{itemize}
\setlength{\itemindent}{-1.5em}
 \item \textbf{Trust-MF \cite{p4}:} this is a recommendation approach proposed by Yang \emph{et al.} in \textit{IJCAI-13}. Trust-MF introduced the relations between users to help improve the recommendation results. In the DBLP dataset used in our experiment, in terms of the relations between users, we use the co-author relationship. For the MeetUp dataset, we use the friendship relation to denote the relation between users in our experiment. As in \cite{p4}, the parameters of this algorithm is set as follows: $\lambda=0.001$ and $\lambda_t=1$. For more details of parameter assignment of Trust-MF please refer to \cite{p4}.
 \item \textbf{Hete-MF \cite{p18}:} this is an approach proposed by Yu \emph{et al}. in \emph{IJCAI-HINA'13}. Hete-MF considers the relations between 'items'. In this paper, for the relations between 'items', we add all the selected relations: $C-P-P-C,~C-P-A-P-C,~C-P-T-P-C$ in DBLP, and $G-U-G,~G-E-L-E-G,~G-U-U-G$ in Meetup.
\end{itemize}

\subsubsection{Evaluation Methods}
As in the following previous recommendation systems research \cite{p4,p18}, we use two evaluation methods in our experiment:

\begin{itemize}
\setlength{\itemindent}{-1.5em}
\item \textbf{MAE:} Mean Absolute Error, which is denoted as follows:
\begin{equation*}
\small
MAE = \frac{1}{T}\sum_{i=0}^{n}\sum_{j=0}^{m}|R_{i,j}-\hat{R_{i,j}}|,
\end{equation*}
where $R_{i,j}$ denotes the ground truth rating that User $i$ gives to Item $j$. $\hat{R_{ij}}$ denotes the rating calculated by the selected algorithm and $T=mn$ is the number of ratings.
\item \textbf{RMSE:} Root Mean Square Error, which is denoted as follows:
\begin{equation*}
\small
RMSE = (\frac{1}{T}\sum_{i=0}^{n}\sum_{j=0}^{m}|R_{i,j}-\hat{R_{i,j}}|^2)^{\frac{1}{2}}
\end{equation*}
The notations in RMSE have the same meaning as in MAE.
\end{itemize}

\subsection{Result and Analysis}
The experimental results are reported in Table \ref{performence}.
From these results, we can see that for our Hete-CF algorithm, with the increase of the training data, its performance becomes better. This is because more training data can provide more information and more importantly, our model can avoid over-fitting.
On the other hand, In the DBLP dataset, we can see that when the dimension of feature vector $d=5$, the performance is always better. This is because our datasets extract nearly five areas of data from the DBLP dataset (see Section \ref{Experiment:data}). Compared with the other algorithms, Hete-CF always performs better. From this aspect, our algorithm has significant advantages when performing the recommendation task on heterogeneous social networks.

\begin{table*}
\caption{Algorithm Performance Comparison in DBLP and MeetUp.}
\label{performence}
\begin{center}

\renewcommand{\arraystretch}{1.2}
\setlength\tabcolsep{3pt}
\small
\begin{tabular} {*{9}{l}}
\hline\noalign{\smallskip}
\textbf{\%Training} & \textbf{Feature} & \textbf{Evaluation} & \textbf{UserMean} & \textbf{ItemMean} & \textbf{NMF} & \textbf{Trust-MF} & \textbf{Hete-MF} & \textbf{Hete-CF}\\
\noalign{\smallskip}
\hline
\noalign{\smallskip}
\textbf{DBLP Dataset} &~&~&~&~&~&~&~&~\\
\multirow{4}*{\centering{$40 \%$}} & \multirow{2}*{$d=5$}
	& MAE & $0.942\pm 0.02$ & $1.065\pm0.02$ & $2.156\pm0.02$ & $0.831\pm0.01$ & $0.931\pm0.02$ & $\textbf{0.831}\pm\textbf{0.02}$ \\
\cline{3-9}
	&~& RMSE & $1.216\pm0.01$ & $1.123\pm0.02$ & $2.394\pm0.01$ & $1.013\pm0.02$ & $1.105\pm0.01$ & $\textbf{1.002}\pm\textbf{0.03}$ \\
\cline{2-9}
& \multirow{2}*{$d=10$}
	& MAE & $0.943\pm0.03$ & $0.948\pm0.01$ & $2.194\pm0.03$ & $0.887\pm0.01$ & $0.901\pm0.01$ & $\textbf{0.859}\pm\textbf{0.01}$ \\
\cline{3-9}
	&~& RMSE & $1.138\pm0.02$ & $1.256\pm0.04$ & $2.292\pm0.02$ & $1.083\pm0.03$ & $1.114\pm0.03$ & $\textbf{1.056}\pm\textbf{0.02}$ \\
\hline

\multirow{4}*{\centering{$60 \%$}} & \multirow{2}*{$d=5$}
	& MAE & $0.948\pm0.02$ & $0.919\pm0.01$ & $2.131\pm0.04$ & $\textbf{0.812}\pm\textbf{0.02}$ & $0.891\pm0.02$ & $0.831\pm0.02$ \\
\cline{3-9}
	&~& RMSE & $1.132\pm0.02$ & $1.157\pm0.01$ & $2.385\pm0.01$ & $\textbf{0.907}\pm\textbf{0.02}$ & $1.010\pm0.03$ & $0.938\pm0.02$ \\
\cline{2-9}
& \multirow{2}*{$d=10$}
	& MAE & $0.932\pm0.03$ & $0.978\pm0.03$ & $2.184\pm0.02$ & $0.873\pm0.03$ & $0.881\pm0.01$ & $\textbf{0.856}\pm\textbf{0.02}$ \\
\cline{3-9}
	&~& RMSE & $1.154\pm0.02$ & $1.143\pm0.02$ & $2.275\pm0.01$ & $1.051\pm0.01$ & $1.013\pm0.02$ & $\textbf{0.994}\pm\textbf{0.03}$ \\
\hline

\textbf{MeetUp Dataset} &~&~&~&~&~&~&~&~\\
\multirow{4}*{\centering{$40 \%$}} & \multirow{2}*{$d=5$}
	& MAE & $1.076\pm0.01$ & $0.954\pm0.02$ & $1.154\pm0.03$ & $0.887\pm0.02$ & $1.035\pm0.02$ & $\textbf{0.854}\pm\textbf{0.01}$ \\
\cline{3-9}
	&~& RMSE & $1.154\pm0.02$ & $1.143\pm0.03$ & $1.354\pm0.03$ & $1.046\pm0.02$ & $1.154\pm0.02$ & $\textbf{0.986}\pm\textbf{0.02}$ \\
\cline{2-9}
& \multirow{2}*{$d=10$}
	& MAE & $0.943\pm0.03$ & $1.065\pm0.01$ & $1.154\pm0.03$ & $0.876\pm0.03$ & $1.013\pm0.03$ & $\textbf{0.854}\pm\textbf{0.01}$ \\
\cline{3-9}
	&~& RMSE & $1.121\pm0.04$ & $1.265\pm0.03$ & $1.264\pm0.02$ & $1.076\pm0.01$ & $1.235\pm0.01$ & $\textbf{0.967}\pm\textbf{0.02}$ \\
\hline

\multirow{4}*{\centering{$60 \%$}} & \multirow{2}*{$d=5$}
	& MAE & $0.976\pm0.02$ & $1.054\pm0.02$ & $1.154\pm0.01$ & $0.854\pm0.01$ & $0.964\pm0.02$ & $\textbf{0.837}\pm\textbf{0.03}$ \\
\cline{3-9}
	&~& RMSE & $1.167\pm0.01$ & $1.136\pm0.01$ & $1.343\pm0.02$ & $1.054\pm0.04$ & $1.132\pm0.01$ & $\textbf{0.966}\pm\textbf{0.02}$ \\
\cline{2-9}
& \multirow{2}*{$d=10$}
	& MAE & $0.965\pm0.02$ & $0.943\pm0.02$ & $1.112\pm0.02$ & $0.887\pm0.02$ & $0.953\pm0.03$ & $\textbf{0.876}\pm\textbf{0.01}$ \\
\cline{3-9}
	&~& RMSE & $1.254\pm0.01$ & $1.254\pm0.02$ & $1.254\pm0.02$ & $1.052\pm0.01$ & $1.012\pm0.02$ & $\textbf{0.978}\pm\textbf{0.02}$ \\
\hline

\end{tabular}
\end{center}
\end{table*}

\subsection{Efficiency Study}
\label{Experiment:eff}
In this section we study the efficiency of our algorithm. As in \cite{p19}, we use the synthetic datasets considering the fact that we can manipulate the size of the dataset flexibly.

\begin{figure}[t]
\label{fig:4}
\centering
\subfigure[Varying Feature Dimension]
{
\includegraphics[width=1.5in]{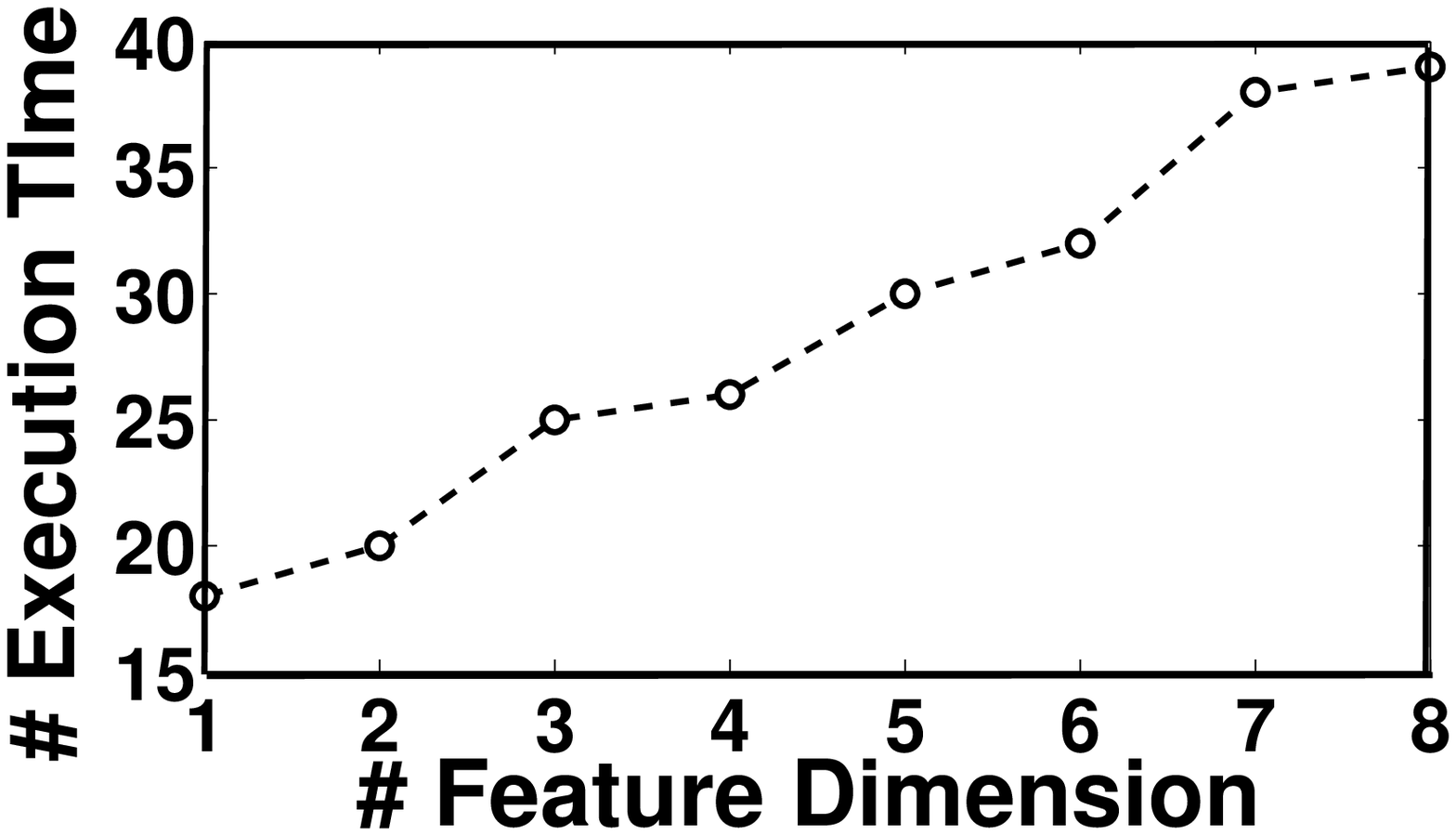}
\label{fig:4:a}
}
\subfigure[Varying Data Size]
{
\includegraphics[width=1.5in]{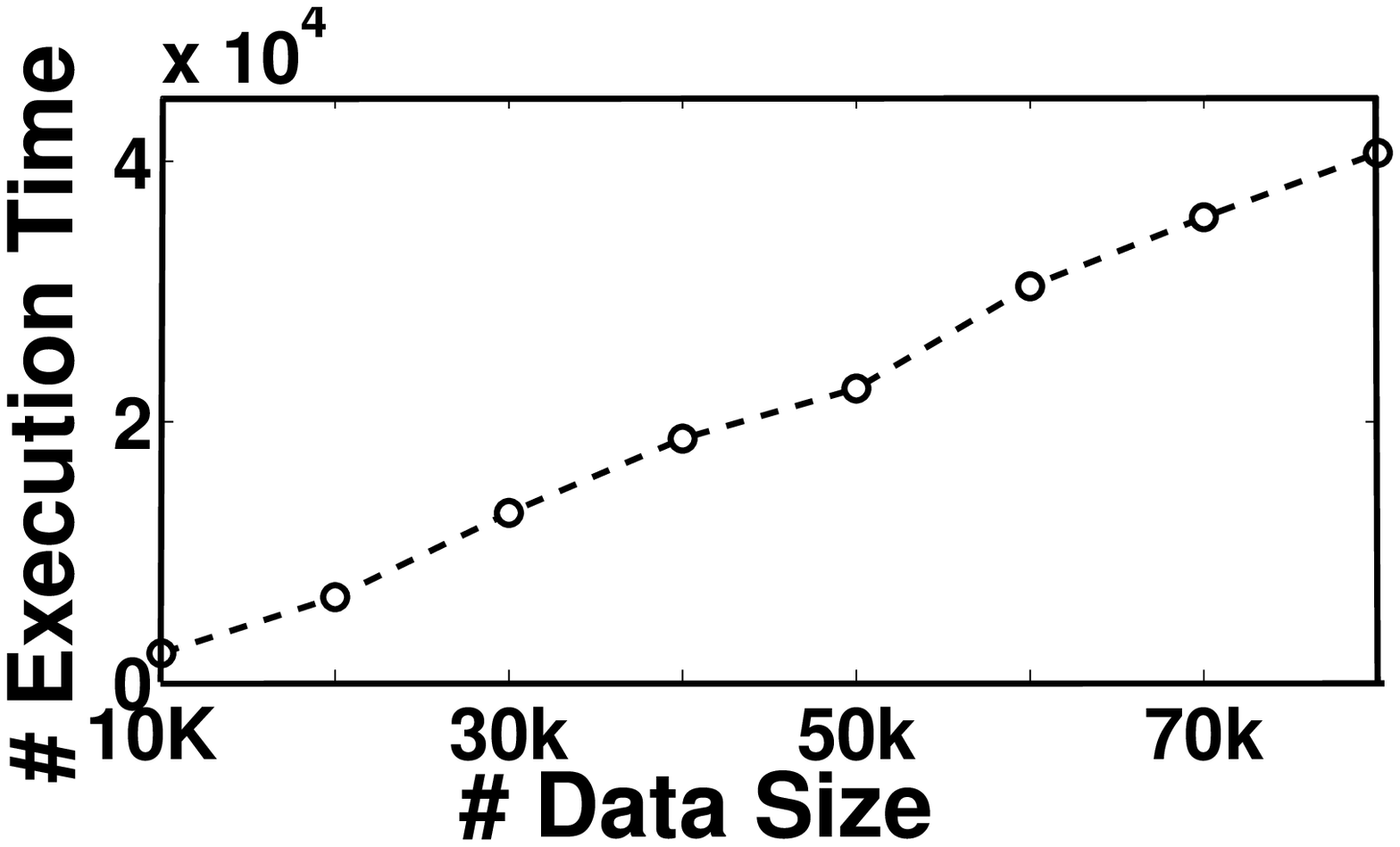}
\label{fig:4:b}
}
\caption{Efficient Study, in (a) we fix the size of dataset, and vary the number of feature dimensions; in (b) we fix the feature dimension $(d = 10)$ and vary the datasets.}
\end{figure}

In Fig.\ref{fig:4:a}, we fix the size of data, and vary the number of the feature dimensions. From the results we can see that the running time of Hete-CF is almost linear to the feature dimension. In Fig. \ref{fig:4:b}, we keep the number of feature dimensions $(d=10)$ fixed and randomly link 2$0\%$ of the objects in the datasets. Then we vary the data size. We can see that the CPU execution time increases with the increase of the data size, but the increase of CPU time is linear to the data size.

\subsection{Parameter Study}
\label{Experiment:para}

\begin{figure}[t]
\centering
\subfigure[DBLP Dataset]
{\includegraphics[width=1.5in]{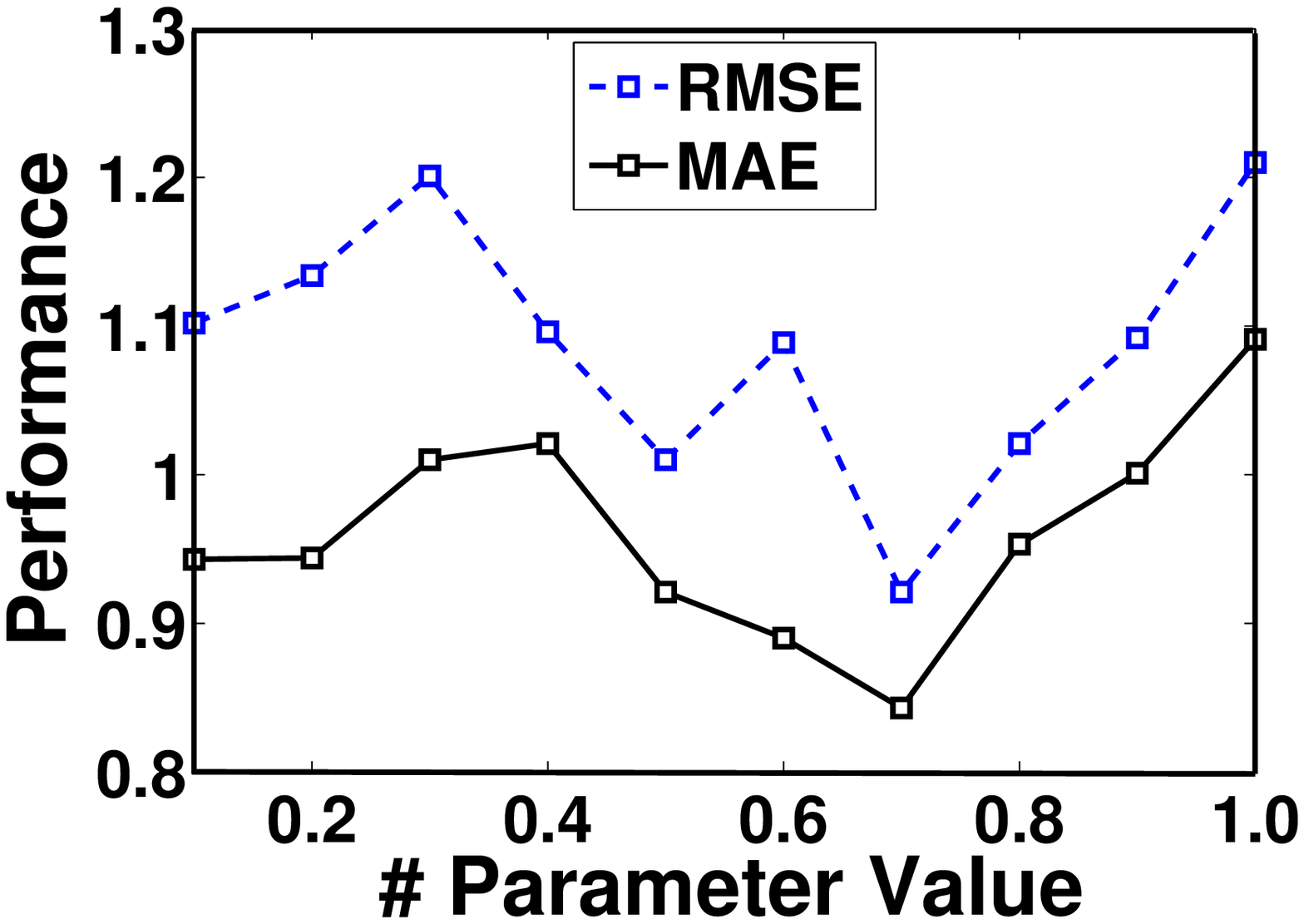}
\label{parameter:dblp}}
\subfigure[MeetUp Dataset]
{\includegraphics[width=1.5in]{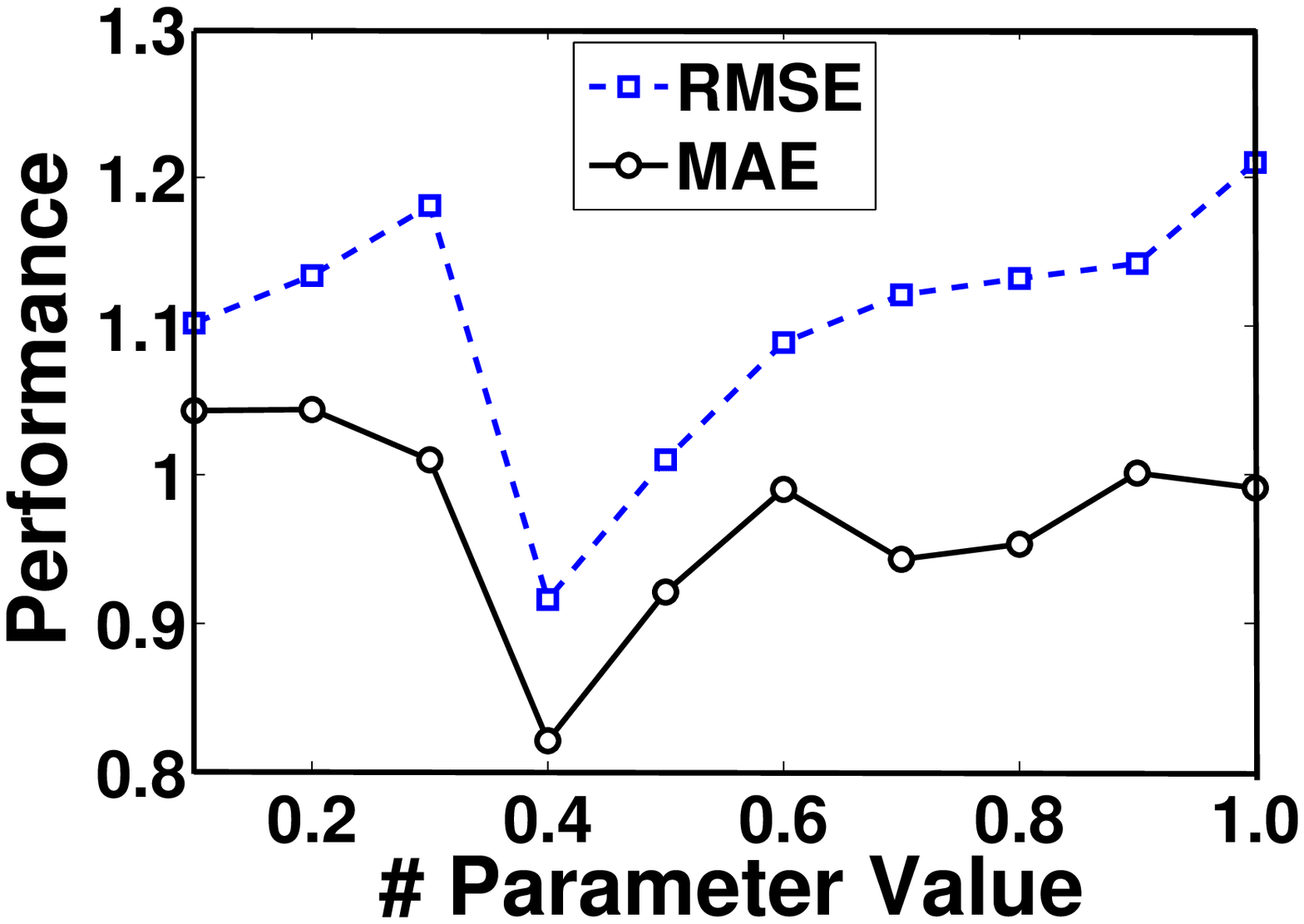}
\label{parameter:meetup}}
\caption{Parameter investigation on the DBLP and MeetUp datasets, $\mu$ ranges from $0.1$ to $1.0$. The value of $\mu$ calculated by our method is: $\mu=0.7$ in DBLP and $\mu=0.4$ in MeetUp}
\label{parameter}
\end{figure}

In this section, we study the impact of parameter $\mu$ in Equation (\ref{model}), which is used for leveraging the importance of the introduced meta-paths between users and items in our model. The result is shown in Fig. \ref{parameter}.

In Fig. \ref{parameter}, we can see that when the value of $\mu$ is too small or too large, the result will not be good enough. Because $\mu$ is used for leveraging the \emph{user-item} meta-path term (the fourth term) in our model (Equation (\ref{model})), the extreme value of $\mu$ (too large or too small) will bias the result.

On the other hand, the $\mu$ value calculated by Equation (\ref{paraset}) is $\mu = 0.7$ in DBLP and $\mu = 0.4$ in MeetUp. And the best result just appeared when $\mu=0.7$ and $\mu=0.4$, as shown in Fig. \ref{parameter}. This demonstrates that our proposed method of evaluating the $\mu$ value is effective, and it can deal with the parameter pre-assignment issue.

\subsection{Case Study for Meta-path Weight}
\label{Experiment:weight}

The output of our algorithm not only contains the predicted ratings but also the weights for each selected meta-path. In this section, we study the learned weight for each meta path. The weight for each selected meta-path is shown in Table \ref{dblpweight} and Table \ref{meetupweight}. For the ease of illustration, we mapped all the weights for each meta-path to the range $[0,1]$.

From the result in Table \ref{dblpweight}, we can see that the meta-path ``$A-P-A$" has a bigger weight, which means the co-author relationship can make an important impact on conference recommendation. The meta-path ``$C-P-A-P-C$" also has a big weight, that is because the meta-path ``$C-P-A-P-C$" contains the target recommendation meta-path ($A-P-C$) and may have a high impact on the conference recommendation.
The meta-path ``$A-P-P-C$" has a high weight too, which is consistent with the ground truth that authors often submit their papers to the conference which they often cites.

From the result in Table \ref{meetupweight}, we can see that the meta-path ``$U-U$" has a bigger weight. This means the friendship between users has a high impact on Group recommendation, which is consistent with human intuition. The meta-path ``$G-U-G$" also has a big weight, that is because $G-U-G$ contains the target recommendation relation ``$U-G$". The Meta-Path ``$U-E-G$" has high weight, which means if an user has participated an Event of a Group, the user may have high probability to be a member of that Group.



\begin{table}
\caption{\emph{Meta-path} Weight of DBLP}
\label{dblpweight}
\begin{center}

\renewcommand{\arraystretch}{1.1}
\setlength\tabcolsep{4pt}
\small
\begin{tabular} {l|l|l}
\hline
Meta-path Type & Meta-path & Weight \\
\hline
\multirow{3}*{\centering{Author$-$Author}}
& $A-P-A$ & $0.4\sim0.5$ \\
\cline{2-3}
	& $A-P-C-P-A$ & $0.3\sim0.4$ \\
\cline{2-3}
	& $A-P-T-P-A$ & $0.2\sim0.3$ \\
\hline
\multirow{3}*{\centering{Conf.$-$ Conf.}}
& $C-P-P-C$ & $0.2\sim0.3$ \\
\cline{2-3}
	& $C-P-A-P-C$ & $0.5\sim 0.6$ \\
\cline{2-3}
	& $C-P-T-P-A$ & $0.2\sim0.3$ \\
\hline
\multirow{2}*{\centering{Author$-$ Conf.}}
& $A-P-P-C$ & $0.7\sim 0.8$ \\
\cline{2-3}
	& $A-P-T-P-C$ & $0.2\sim 0.3$ \\
\hline

\end{tabular}
\end{center}
\end{table}

\begin{table}
\small
\caption{\emph{Meta-path }Weight of MeetUp}
\label{meetupweight}
\begin{center}

\renewcommand{\arraystretch}{1.1}
\setlength\tabcolsep{4pt}
\small
\begin{tabular} {l|l|l}
\hline
Meta-path Type & Meta-path & Weight \\
\hline
\multirow{4}*{\centering{User$-$User}}
	& $U-U$ & $0.4\sim 0.5$ \\
\cline{2-3}
& $U-L-U$ & $0.1\sim0.2$ \\
\cline{2-3}
	& $U-G-U$ & $0.3\sim0.4$ \\
\cline{2-3}
	& $U-E-U$ & $0.1\sim0.2$ \\
\hline
\multirow{3}*{\centering{Group$-$Group}}
& $G-U-G$ & $0.4\sim0.5$ \\
\cline{2-3}
	& $G-E-L-E-G$ & $0.2\sim0.3$ \\
\cline{2-3}
	& $G-U-U-G$ & $0.3\sim0.4$ \\
\hline
\multirow{3}*{\centering{User$-$Group}}
& $U-U-G$ & $0.3\sim0.4$ \\
\cline{2-3}
	& $U-E-G$ & $0.4\sim0.5$ \\
\cline{2-3}
	& $U-L-E-G$ & $0.2\sim0.3$ \\
\hline

\end{tabular}
\end{center}
\end{table}
\section{Related Work}
\label{related}
In this section, we introduce some related works to our research.

\subsection{Collaborative Filtering and Social-based Recommendation}
\label{related:cf}

Collaborative Filtering Recommendation is a popular method, which has been researched intensively in recent years. In \cite{p4}, the authors introduced the social trust information and proposed a novel collaborative filtering model. In \cite{p15}, the authors proposed a parallel collaborating algorithm for the Netflix Prize, Alternating-Least-Squares with Weighted-¦Ë-Regularization (ALS-WR). In \cite{p17}, the authors proposed a probabilistic matrix factorization model, which can deal with problems with large-scale datasets and sparse ratings. In this research, we have incorporated the ideas presented in \cite{p15} and \cite{p17} to improve our model.

Social based recommendation is an emerging research topic which combines the recommendation algorithms and social media mining algorithms. In \cite{p2}, the authors proposed a general social based recommendation framework using social regularization. This work demonstrates that the social network information can benefit recommender systems. In \cite{p6}, the author proposed a group recommendation method on EBSN. This method considers location features, social features, and implicit patterns in a unified model. Another recommendation system proposed on EBSN is LCARS \cite{p3}. LCARS consists of an off-line model, which considers the off-line part of EBSN, and an on-line recommendation model. By considering both the on-line and off-line parts, the recommendation performance was significantly improved \cite{p3}. In \cite{p7}, the authors proposed to realise location recommendation services. The recommendation method in \cite{p7} considered both the friend relationship and the geographic information.
The algorithm proposed in \cite{p13} utilised the knowledge from other domains to improve the recommendation performance. In \cite{p5}, the authors proposed a recommendation algorithm on HIN and utilise part of the heterogeneous relationships.

In this research, we aim to bridge the gap between CF-based recommendation and social-based recommendation.
Different from the above algorithms, our proposed method considers the recommendation in HSNs (e.g., LBSN, EBSN and other HINs with social information). In addition, our method utilises all types of relations in HSNs while the above algorithms utilise only part of the relations.

\subsection{Mining Heterogeneous Social Network}
\label{related:HSN}

Heterogeneous Social Network is defined as the heterogeneous information network with social information in our paper. As introduced before, Heterogeneous Social Network is a special case of HIN. HIN (Heterogeneous Information Network) has multi-typed objects and relations, and it may contain more meaningful information. The concept of heterogeneous information network is first proposed by Sun \emph{et. al} in \cite{p10}. The works proposed by them \cite{p3,p5,p10,p12,p16,p18,p19} also demonstrated that by mining heterogeneous information network, one can obtain more meaningful results. It also attracted other researchers to investigate on mining the heterogeneous information network \cite{p22,p23}.

There are also some other kinds of heterogeneous social networks as previously introduced in Section \ref{preliminary:HSN}. Location based networks \cite{p11} are heterogenous social networks with location information, and they have attracted many researchers in recent years \cite{p24,p25}. In \cite{p24}, the authors proposed a model to mine interesting locations using GPS data; in \cite{p25}, the authors designed a model to explain the human movements using some location-based network data. Event based social networks (EBSNs) are those heterogeneous social networks which contain not only social information but also the off-line event information. Recently EBSNs also receive more and more attention \cite{p1,p6}. In \cite{p1}, the authors formally proposed the definition of EBSNs, and some initial research was carried out in \cite{p1} on event based social networks, such as community detection and link prediction.

In this research, we consider all types of heterogeneous social network together, and our recommendation method can be used in all the networks introduced above.

\balance

\section{Conclusion and Future Work}
\label{conclusion}

Collaborative filtering recommendation are frequently suffered from the data sparsity and cold start problems. Social based recommendation, as one of the efficient and emerging methods to deal with these problems, has attracted much attention in recent years. In this paper, we focus on the recommendation problem using social heterogeneous relations, and we proposed Hete-CF, a Collaborative Filtering recommendation method on Heterogeneous Social Networks. Different from the previous social based recommendation methods, we propose to effectively incorporate all the social relations, including the relations between users, items and user-item. In addition, since Hete-CF is a network structure based model, it can be used in many types of social networks (e.g. , LBSN, EBSN, and other HINs with social information). For instance, Hete-CF can be used in recommending off-line events in EBSN or recommending locations (or hotels) in LBSN. The experiments on two real-world heterogeneous social networks (DBLP and Meetup) demonstrate the effectiveness and efficiency of Hete-CF.

In the future, we intend to apply Hete-CF to more real-world recommendation problems. In addition, another direction of our future research is to explore the potential of Hete-CF on big data problems, such as problems involving massive amounts social media data.



\setlength{\bibsep}{1ex}  
\bibliographystyle{IEEEtran}
\bibliography{bibliography}

\end{document}